\documentclass[twocolumn,
showpacs,preprintnumbers,amsmath,amssymb,floatfix,aps]{revtex4}

\usepackage{graphicx}
\usepackage{dcolumn}
\usepackage{bm}

\newcommand{\nrt}{n({\bf r},t)}
\newcommand{\gtwo}{g^{(2)}}
\newcommand{\gud}{g^{(2)}_{\uparrow \downarrow}}

\begin{document}

\preprint{APS/123-QED}

\title{Pair correlations of an expanding superfluid Fermi gas} %

\author{C. Lobo}\email{clobo@science.unitn.it}
\author{I. Carusotto} 
\author{S. Giorgini} 
\author{A. Recati} 
\author{S. Stringari}
\affiliation{Dipartimento di Fisica, Universit\`a di Trento
and CNR-INFM BEC Center, I-38050 Povo, Trento, Italy \\}

\begin{abstract}
The pair correlation function of an expanding gas is investigated 
with an emphasis on the BEC-BCS crossover of a superfluid Fermi gas 
at zero temperature. At unitarity quantum Monte Carlo simulations reveal the 
occurrence of a sizable bunching effect due to interactions in the spin up-down 
channel which, at short distances, is larger than that exhibited by thermal 
bosons in the Hanbury-Brown and Twiss effect. We propose a local 
equilibrium ansatz for the pair correlation function which we predict will 
remain isotropic during the expansion even if the trapping potential is 
anisotropic, contrary to what happens for non-interacting gases. This behavior is 
understood to be a consequence of the violation of scaling of the pair 
correlation function due to interactions. 
\end{abstract}
\pacs{Valid PACS appear here}

\maketitle
Recent experimental studies of two-body correlations in an expanding atomic
cloud are opening new perspectives in the study of quantum-statistical effects 
in ultracold gases. These include the measurement of the Hanbury-Brown and Twiss
effect~\cite{loudon} in a dilute Bose gas, both concerning spatial and temporal 
correlations~\cite{aspect,esslinger}, measurements in the Mott insulating phase 
of ultracold gases in an optical lattice~\cite{bloch}, as well as the study of 
atomic pair correlations in a Fermi gas after dissociation of molecules near a 
Feshbach resonance~\cite{jin}. In some cases the measured quantities are related 
to the real space properties ({\it i.e.} to the pair correlation function), whereas 
in other cases they refer to correlations in momentum space. In all cases, 
measurements are done after a period of free expansion since the gas {\it in situ} 
does not provide sufficient optical resolution.

The pair correlation function is defined as
\begin{equation}
g^{(2)}_{\sigma \sigma^\prime}({\bf r}_1,{\bf r}_2)=
  \frac{n^{(2)}({\bf r}_1,\sigma;{\bf 
  r}_2,\sigma^\prime)}{n({\bf r}_1,\sigma)n({\bf r}_2,\sigma^\prime)} \;,
\end{equation}
where
\begin{equation}
n^{(2)}({\bf r}_1,\sigma;{\bf
  r}_2,\sigma^\prime)=\langle \psi^\dagger_\sigma ({\bf r}_1)
\psi^\dagger_{\sigma^\prime} ({\bf r}_2)\psi_{\sigma^\prime} ({\bf
r}_2)\psi_\sigma ({\bf r}_1) \rangle \;,
\end{equation}
is the second order correlation function and $n({\bf r},\sigma)$ is the density of the 
$\sigma$ spin component.
The spin indices $\sigma$, $\sigma^\prime$ are used if there is more than one 
species of atoms present. The pair correlation function is 
a key quantity in  many-body physics, being sensitive to both statistical and
interaction effects. The statistical effects appear, for example, in non-interacting gases of 
identical bosons above the critical temperature, where they are responsible for a 
characteristic bunching effect which has been recently the object of experimental 
measurements~\cite{aspect}. 

\begin{figure}
\begin{center}
\includegraphics*[width=8.5cm]{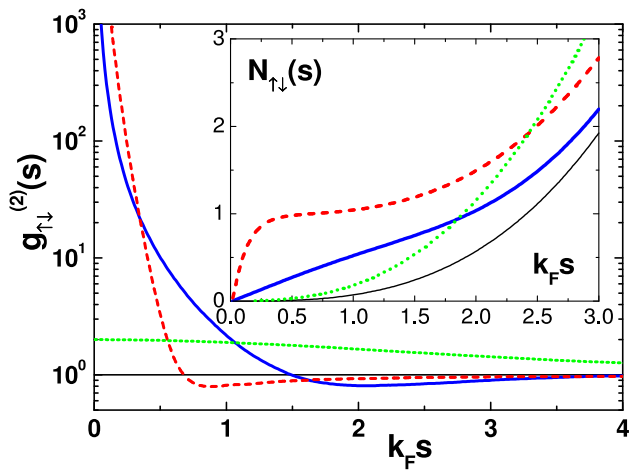}
\caption{(color online). Spin up-down pair correlation function $\gtwo_{\uparrow\downarrow}(s)$ of
a homogeneous system.
Dashed line (red online): Fermi gas in the deep BEC regime ($1/k_Fa=4$); solid line (blue online): 
Fermi gas at unitarity ($1/k_Fa=0$); dotted line (green online): ideal Bose gas with density 
$n=k_F^3/(6\pi^2)$ at the Bose-Einstein critical temperature; thin solid line (black online): 
uncorrelated gas with $\gtwo(r)=1$. Inset: Integrated pair correlation function 
$N_{\uparrow\downarrow}(s)$ for the same configurations.} 
\label{fig1}
\end{center}
\end{figure}

In this Letter we discuss the case of a Fermi gas, with 
equal population of two spin states, close to a Feshbach resonance where the value of the
scattering length $a$ can be tuned by varying the external magnetic field. The opposite spin  
$g^{(2)}_{\uparrow \downarrow}$ correlation function is strongly affected by interactions and 
for a homogeneous system has been calculated using fixed-node diffusion Monte Carlo 
simulations~\cite{stefanoMC}. The result at $T=0$ is shown in Fig.~\ref{fig1} 
where $\gtwo_{\uparrow\downarrow}(s)$ is plotted as a function of the dimensionless variable 
$k_Fs$ for various configurations. In the inset we plot the integrated quantity
$N_{\uparrow\downarrow}(s)= n\int_0^s 4\pi r^2 dr\gtwo_{\uparrow\downarrow}(r)$ giving the 
average number of atoms of opposite spin within a sphere of radius $s$ around a given atom. 
Here $n=k_F^3/(6\pi^2)$ is the single spin density and  $k_F$ is the Fermi momentum. On the 
BEC side of the resonance 
($k_Fa=0.25$, dashed line) there is a pronounced bunching effect at short distances due to the 
presence of diatomic molecules and the pair correlation function rapidly approaches the 
uncorrelated value $\gtwo_{\uparrow\downarrow}(s)=1$ (thin solid line) at distances of the order 
of the scattering length. As a consequence, the integrated probability $N_{\uparrow\downarrow}(s)$ 
reaches 1 at $k_Fs \sim 0.25$ in contrast to the uncorrelated gas where the same value is reached 
much later, at $k_F s=2.4$. In the unitary limit ($1/k_F a=0$, solid line) 
$\gtwo_{\uparrow\downarrow}$ is a universal function of $k_Fs$, the inverse of $k_F$ providing 
the only length scale of the problem. In this case the bunching effect is reduced 
as compared with the BEC case, but is still important (for example at $k_F s=1$ the value of 
$N_{\uparrow\downarrow}(s)$ is 0.5 whereas the non-interacting value is negligible). 
For comparison we plot the pair correlation function for a gas of non-interacting bosons with 
density $n=k_F^3/(6\pi^2)$ at the Bose-Einstein critical temperature $T_c$ (dotted line). At short 
distances $\gtwo(s)=2$ and the corresponding integrated probability $N(s)$, while larger than that 
of the uncorrelated gas, is still quite small compared to the  unitary case~\cite{interactions}. 
The sizable effect 
exhibited by the $\gtwo_{\uparrow\downarrow}(s)$ correlation function in interacting Fermi gases 
is due to the presence of correlations resulting in a $1/s^2$ behavior as $s\to 0$~\cite{notes0} 
which characterizes not only the deep BEC limit, but the whole 
crossover~\cite{noteupup}. The short range behavior of $\gtwo_{\uparrow\downarrow}$
has recently been the subject of experimental studies using spectroscopic techniques~\cite{jinhulet}.  
At unitarity, the $s\to 0$ limit of $s^2\gtwo_{\uparrow\downarrow}(s)$ is fixed by many-body effects 
and the quantum Monte Carlo simulation yields the value $(k_Fs)^2\gtwo_{\uparrow\downarrow}(s)\to 2.7$. 
In the BEC limit one finds instead the value $(k_Fs)^2\gtwo_{\uparrow\downarrow}(s)\to 3\pi/(k_Fa)$ 
characterizing a free molecule.

As already mentioned it is difficult to directly measure the pair correlation function in real space
while the atomic gas is confined in the trap since the cloud is too small to optically resolve 
features at interatomic distances~\cite{measurementSq}. The gas needs to be released, 
undergoing an expansion,
before any measurement can be made. If the scattering length is kept constant during the expansion, 
the calculation of the time dependent evolution of the pair correlation function is strongly affected 
by interactions and represents a challenging theoretical problem for which few exact solutions are known. 
One of them is the unitary Fermi gas released from an isotropic
harmonic potential~\cite{castin,tonks}. In the more general case (including anisotropic 
traps or systems which are not at unitarity) the behavior of the expansion can 
be nevertheless 
predicted if the system is in the hydrodynamic (HD) regime, corresponding to a local equilibrium 
condition. In this case one can use the ansatz
\begin{equation}
g^{(2)}({\bf r}_1,{\bf r}_2,t)= g^{(2)}_\textrm{hom}({\bf s},n({\bf r},t)) \;, 
\label{eq:hdg2} 
\end{equation}
for the pair correlation function, based on a local density approximation. Here 
${\bf r}=({\bf r}_1+{\bf r}_2)/2$ and ${\bf s}={\bf r}_1-{\bf r}_2$ are, respectively, the center of 
mass and relative coordinates and $g^{(2)}_\textrm{hom}$ is the pair correlation function for a 
homogeneous gas with density $n({\bf r},t)$. From an experimental point of view the ansatz 
(\ref{eq:hdg2}) relates the locally measured density $n({\bf r},t)$ to the locally measured 
$\gtwo$. From a theoretical perspective it gives us the value of $\gtwo$ during the expansion if we 
know the equilibrium form of $g^{(2)}_\textrm{hom}(s)$ (from, e.g., quantum Monte Carlo calculations) 
and the time evolution of the local density $n({\bf r},t)$. In particular, at unitarity, where 
$\gtwo_{hom}$ depends only on the combination $k_Fs$, the expansion acts like a microscope, the value 
of $k_F\propto n^{1/3}$ being reduced as a function of time.

The dynamics of the density $n({\bf r},t)$ can be analysed employing the HD theory. This theory has been 
quite successful in predicting the behavior of $n({\bf r},t)$ in several important configurations. For 
example in the case of a dilute Bose gas the hydrodynamic predictions for the time evolution of the 
density agree in an excellent way with the exact solutions of the Gross-Pitaevskii equation and with 
experiments (see, for example, Ref.~\cite{rmp99}). They have been used to study, among others, the collective 
oscillations~\cite{sandro96} and the expansion after release from the trap~\cite{castindum,kagan,menotti}. 
If the trapping has the harmonic form $V({\bf r})=\sum_{i=x,y,z} m\omega_i^2r_i^2/2$ and the chemical 
potential follows a power law dependence on the density, $\mu(n)\propto n^{\gamma}$ the HD predictions 
for the expanding density profile can  be written in the scaled form
\begin{equation}
n({\bf r},t)=\frac{1}{b_xb_yb_z}n_0(x/b_x,y/b_y,z/b_z)\;,
\label{nHD}
\end{equation}
with the scaling coefficients $b_i$ obeying  the system~\cite{castindum}
\begin{equation}
\ddot{b}_i-\frac{\omega_i^2}{b_i(b_xb_yb_z)^\gamma}=0\;,
\label{HDeqs}
\end{equation}
of nonlinear coupled equations with the initial conditions $b_i(t=0)=1$ and $n_0({\bf r})\equiv n({\bf r},t=0)$.
The polytropic dependence of the chemical potential applies both to the case of a dilute Bose gas 
($\gamma=1$) and to the unitary limit of a Fermi gas ($\gamma=2/3$). When released from an anisotropic trap, 
the aspect ratio of the expanding density profile exhibits an inversion from the cigar to the disc geometry 
(and viceversa) as a consequence of the HD forces which are more active along the directions
where the gradient of the density is larger. Therefore the density profile at long times will be 
anisotropic. On the other hand we see from Eq. (\ref{eq:hdg2}) that the pair correlation function will
keep its isotropy as a function of ${\bf s}$. 

The ansatz (\ref{eq:hdg2}), which assumes local equilibrium for the description of the expansion, requires more 
stringent conditions compared to the usual local density approximation applied to stationary configurations. 
In the latter case, the only relevant condition is that the typical length scale, where $g^{(2)}$ approaches 
the uncorrelated value $g^{(2)}=1$, be much smaller than the size of the cloud. The use of Eq. (\ref{eq:hdg2}),
instead, also requires that adiabaticity be ensured during the expansion. In the superfluid regime, 
HD theory provides a justification for the local equilibrium approximation. We might however wonder 
whether superfluidity is preserved during the expansion. In a Fermi gas at $T=0$, the disappearance of 
superfluidity could take place through pair breaking mechanisms. 
Since a large number of quantized vortices are observed after expansion, recent experiments~\cite{mitmartin} 
suggest that the system remains superfluid during the expansion on the BEC side of the crossover. The robustness 
of superfluidity is also suggested by the 
fact that the gap is much larger than the typical oscillator frequency (whose inverse is the timescale of the 
expansion) and that, as a consequence, pairs cannot break during the expansion. This argument 
holds also at unitarity where the gap is of the order of the Fermi energy. On the other hand, superfluidity 
is more fragile on the BCS side because the gap becomes exponentially small during the expansion 
(for a measurement of the pairing gap in ultracold Fermi gases along the BEC-BCS crossover see 
Ref.~\cite{grimm}).

The experimental measurement of the pair correlation function and the verification of Eq. (\ref{eq:hdg2}) requires 
that certain conditions be met. In particular, the expansion time should be long enough for features at the 
interatomic level to be resolved. By choosing, for example, a trapped gas with central density equal to 
$n\simeq 10^{13}$~cm$^{-3}$ (corresponding to an inverse Fermi momentum equal to $k_F^{-1}\simeq 100$~nm) and 
isotropic trapping, we find that, after a time equal to $\omega t\simeq 20$, the 
density is reduced to a value $\sim 10^{9}$~cm$^{-3}$ corresponding to $k_F^{-1}\sim$~3$\mu$m. This value is 
compatible with the 
present limits of optical resolution~\cite{noteresolut}. At the same time, if we choose a large value of the 
scattering length the system will keep the condition of unitarity $k_Fa\gg 1$ also at the observation time. 

The HD description of the expansion points out the occurrence of important qualitative differences in 
the expansion of interacting and non-interacting gases (see Table~\ref{tab1}). Contrary to what happens in the 
HD regime, when released from an anisotropic trap the density of a non-interacting gas (e.g., fermions or 
uncondensed thermal bosons) 
becomes isotropic at long times during the expansion, reflecting the isotropy of the initial momentum 
distribution~\cite{noteIG}. The pair correlation function 
also exhibits drastic differences. 
In fact, in a non-interacting gas the dependence of $\gtwo$ on the relative coordinate ${\bf s}$ will undergo 
an anisotropic expansion. 
For example, for non-interacting bosonic or fermionic gases trapped in a harmonic potential at 
temperatures much higher than that of quantum degeneracy, the pair correlation function can be shown to exhibit 
the following behavior~\cite{collisional}
\begin{equation}
\gtwo({\bf s},t) = 1\pm \exp\left( -2 \pi \sum_{i=x,y,z} \frac{s^2_i}{\lambda_T^{2}(\omega^2_it^2+1)}\right) \;,
\label{eq:g2ideal}
\end{equation}
where $\lambda_T=\sqrt{2 \pi \hbar^2/m k_B T}$ is the thermal de Broglie wavelength. Here the plus sign is for 
bosons, revealing a bunching effect (as discussed previously), while the minus sign is for fermions which exhibit 
anti-bunching as a consequence of the Pauli principle. For large times, the typical value of the decay length of 
these statistical correlations depends on direction and scales as $\lambda_T \omega_i t=\sqrt{4\pi}\hbar 
t/(m R_i)$, a value that has recently been used to analyze the data in the experiment of~\cite{aspect} with 
thermal bosons. Here $R_i=\sqrt{2k_BT/m\omega_i^2}$ is the {\it in situ} radius of the thermal cloud in the 
$i$-th direction.

\begin{table}
\centering
\caption{Role of anisotropy in the expansion of the density and of the pair correlation function 
of a trapped gas.}
\begin{tabular}{|l|c|c|} \hline
 Behavior after expansion & $\nrt$ & $\gtwo({\bf s},t)$ \\ \hline
Ideal gas & isotropic & anisotropic \\
(bosons $T>T_c$ or fermions) & & \\
 & & \\ \hline
Hydrodynamic regime & anisotropic & isotropic \\
 & & \\ \hline  
\end{tabular}
\label{tab1}
\end{table}

Let us finally discuss the problem of the expansion from the general point of view of scaling, characterized 
by a coordinate transformation of the type:
\begin{equation}
x\to x/b_x(t), \; y\to y/b_y(t), \; z\to z/b_z(t) \;.
\label{eq:xyz}
\end{equation}
An exact scaling solution is known to be obeyed by the many-body wave function of a unitary Fermi gas 
released from an isotropic harmonic 
potential~\cite{castin}. In this case all correlation functions exhibit the same scaling 
behavior and one finds $n({\bf r},t) = n_0({\bf r}/b)/b^3$ and  
$n^{(2)}({\bf r}_1,{\bf r}_2,t) = n^{(2)}({\bf r}_1/b,{\bf r}_2/b,0)/b^6$, with $b=\sqrt{\omega^2t^2+1}$.
The HD picture reproduces this behavior, in fact 
Eqs. (\ref{nHD})-(\ref{HDeqs}), with $\omega_i\equiv\omega$, $b_i(t)\equiv b(t)$ and 
$\gamma=2/3$, yield the same time dependence for the density profile. Agreement is also found for
the time dependence of the pair correlation function. In fact at unitarity the correlation function 
evaluated for a homogeneous medium depends only on the combination $k_Fs\propto n^{1/3}s$ so that 
Eq. (\ref{eq:hdg2}) can be written in the scaled form
\begin{equation}
\gtwo({\bf r}_1,{\bf r}_2,t)=\gtwo_\textrm{hom}\left( n_0^{1/3}({\bf r}/b)s/b \right) \;,
\label{g2}
\end{equation}
which, for isotropic harmonic trapping, agrees with the scaling result $g^{(2)}({\bf r}_1,{\bf r}_2,t) = 
g^{(2)}({\bf r}_1/b,{\bf r}_2/b,0)$. Naturally the agreement holds only for values of $s$ much smaller than 
the size of the system where the local density approximation can be properly applied. 

The applicability of scaling to describe the expansion of the pair 
correlation function is however not general. In fact for an 
interacting gas the boundary conditions on the 
many-body wave function impose restrictions which are in general incompatible with scaling. At distances much 
shorter than the interparticle separation, but much larger than the effective range of interatomic interactions, 
the wave function is determined up to a proportionality constant by the Bethe-Peierls boundary condition: for any 
value of the scattering length $a$ and for all pairs of atoms $ij$, the 
many-body wave function $\Psi$ obeys the condition:
\begin{equation}
\Psi(|{\bf r}_i-{\bf r}_j|\rightarrow 0) \propto \frac{1}{|{\bf r}_i-{\bf r}_j|} - \frac{1}{a}\;, 
\label{eq:boundary}
\end{equation}
where the limit is taken fixing the position of all other atoms as well as that of the center of mass 
$({\bf r}_i+{\bf r}_j)/2$. This boundary condition has two characteristic features: I) it is isotropic and II) it 
introduces a length scale $a$. Any scaling solution of the form of Eq. (\ref{eq:xyz}) will run into two kinds of 
difficulties. I) requires that all the $b_i(t)$ be identical, which is in general incompatible 
with the free 
expansion from an anisotropic trap. II) means that, even if the trap is isotropic, leading to identical $b_i(t)$, 
a rescaling of the coordinates would change the ratio $a/|{\bf r}_i-{\bf r}_j|$ which is required to be fixed. These 
problems do not appear when $a=0$ or $a\rightarrow \infty$, i.e., in the non-interacting and isotropic unitary 
gas~\cite{engineering}. The above argument applies to $\gtwo$ and to all higher order correlation functions as they
all obey the boundary condition (\ref{eq:boundary}). However, it does not affect the density $n({\bf r},t)$ for 
which scaling laws do exist, e.g., for HD systems. 

From I) we can now better understand the behaviour of $\gtwo$ summarised in the Table. It implies that $\gtwo$ must 
remain 
isotropic (at least for small $|{\bf r}_1-{\bf r}_2|$) for all interacting gases whereas in the non-interacting 
limit this constraint is removed and $\gtwo$ is allowed to exhibit the anisotropic scaling of Eq. (\ref{eq:g2ideal}). 
On the other hand, II) leads to a violation of scaling which dramatically illustrates the difference between the 
BEC and unitary regimes: the latter has perfect scaling of $\gud$ if the gas is isotropic. Conversely, in the molecular 
gas in the deep BEC regime, although the density $n({\bf r},t)$ scales during the expansion as in ordinary 
Bose-Einstein condensates, $\gud$ does not since it now contains information about the molecular bound state whose 
size is fixed by the scattering length $a$ and is independent of the local value of the density.

In conclusion, we investigated the spin up-down pair correlation function of a Fermi gas at $T=0$ in the BEC-BCS 
crossover. Under the condition that local equilibrium is ensured during the expansion, we addressed the non trivial 
problem of the time dependence of the correlation function after release of the gas from the confining potential. 
In contrast to the behavior of non-interacting gases, the pair correlation function violates the 
scaling solution (\ref{eq:xyz}) and is predicted to remain locally isotropic during the expansion even for 
anisotropic trapping. At unitarity, the absence of other relevant length scales beside $k_F^{-1}$ makes the 
expansion act like a magnification lens as the average distance between particles becomes visible. 

Acknowledgements: We gratefully acknowledge useful discussions with L.P. Pitaevskii and M.W. Zwierlein.
We also acknowledge support by the Ministero dell'Istruzione, dell'Universit\`a e della Ricerca (MIUR).


\begin{thebibliography}{99}
\bibitem{notes0} Notice that $s\to 0$ here means $R_0\ll s\ll min(a,k_F^{-1})$, where $R_0$ is the effective
range of interatomic forces.
\bibitem{noteupup}  The spin up-up correlation function $\gtwo_{\uparrow\uparrow}$ is less affected by 
interactions than $\gtwo_{\uparrow\downarrow}$. For example, at unitarity, it almost coincides with the 
prediction for a non-interacting Fermi gas~\cite{stefanoMC}.
\bibitem{tonks} Another non trivial case where an exact solution is available is the 1D Tonks gas initially trapped 
by a harmonic potential. In this case $\gtwo$ coincides with the one of the non-interacting Fermi gas and evolves 
in time with a characteristic anti-bunching behavior at short distances.
\bibitem{noteresolut} Notice that the quantity $N_{\uparrow\downarrow}(s)$ plotted in Fig.~\ref{fig1} is
related to $g_{\uparrow\downarrow}^{(2)}(s)$ convoluted with a response function which is flat over a 
sphere of radius $s$ and zero outside.
\bibitem{noteIG} A non-interacting Bose gas at $T=0$ behaves differently, as the
expansion after release is anisotropic. However, in this case, $\gtwo(s)=1$ at any time.
\end{thebibliography}
\end{document}